\documentclass[epj]{svjour}

\usepackage{epsfig}
\usepackage{amsmath}
\usepackage[usenames]{color}
\usepackage{ulem} %% for strike-through

\newcommand{\be}{\begin{eqnarray}}
\newcommand{\ee}{\end{eqnarray}}
\newcommand{\bc}{\begin{center}}
\newcommand{\ec}{\end{center}}
\newcommand{\non}{\nonumber \\}

\begin{document}

\title{Evaluation of the polarization observables
\boldmath$I^S$ and \boldmath$I^C$  in the reaction $\boldmath{\vec{\gamma} 
p \to \vec{\pi}^0 \vec{\eta} p}$}

\author{M. D\"oring\inst{1}\and E.~Oset\inst{2}\and Ulf-G.~Mei{\ss}ner\inst{1,3,4}}
\institute{
Institut f\"ur Kernphysik and J\"ulich Center for Hadron Physics, Forschungszentrum J\"ulich, D-52425 J\"ulich, Germany \thanks{\email{m.doering@fz-juelich.de}} 
\and Departamento de F\'{\i}sica Te\'orica and IFIC, Centro Mixto Universidad de Valencia-CSIC, Institutos de Investigaci\'on de Paterna, Aptdo. 22085, 46071 Valencia, Spain \thanks{\email{oset@ific.uv.es}}
\and Institute for Advanced Simulation, Forschungszentrum J\"ulich, D-52425 J\"ulich, Germany
\and Helmholtz-Institut f\"ur Strahlen- und Kernphysik (Theorie) and Bethe
Center for Theoretical Physics, Universit\"at Bonn, Nu\ss allee 14-16, D-53115
Bonn, Germany \thanks{\email{meissner@hiskp.uni-bonn.de}}
}

\date{ FZJ-IKP-TH-2010-6, HISKP-TH-10/07}

\abstract{
We evaluate the polarization observables $I^S$ and $I^C$ for the reaction
$\gamma p\to\pi^0\eta p$, using a chiral unitary framework developed earlier. 
The $I^S$ and $I^C$ observables have been recently measured for the first time 
by the CBELSA/TAPS collaboration. The theoretical predictions of $I^S$ and
$I^C$, given for altogether 18 angle dependent functions, are in good
agreement with the measurements. Also, the asymmetry $d\Sigma/d\cos\theta$ 
evaluated here agrees with the data. We show the importance of the 
$\Delta(1700)\,D_{33}$ resonance and its $S$-wave decay into 
$\eta\Delta(1232)$. The result can be considered as a further confirmation of 
the dynamical nature of this resonance. At the highest energies, deviations of 
the predictions from the data start to become noticeable, which leaves room
for additional processes and resonances such as a $\Delta(1940)\,D_{33}$.
We also point out how to further improve the calculation.}

\PACS{
      {25.20.Lj}{Photoproduction reactions}\and
      {13.60.Le}{Meson production}\and
      {13.75.Gx}{Pion-baryon interactions}\and
      {14.20.Gk}{Baryon resonances with S=0}
      } 
\maketitle

\section{Introduction}
  
The photoproduction of meson pairs is proving to be a rich field allowing us to widen our understanding of hadron dynamics and  hadron structure. Following   much work devoted to the photoproduction of two pions in the last decade,  both experimental~\cite{Braghieri:1994rf,Zabrodin:1997xd,Harter:1997jq,Zabrodin:1999sq,Wolf:2000qt,Langgartner:2001sg,Assafiri:2003mv,Ripani:2002ss,Ahrens:2003na,Kotulla:2003cx,Ahrens:2005ia,Strauch:2005cs,Ahrens:2007zzj,Thoma:2007bm,:2007bk,:2009fq,Krambrich:2009te} and theoretical \cite{GomezTejedor:1993bq,GomezTejedor:1995pe,Bernard:1994ds,Bernard:1996ft,Bernard:1995tf,Nacher:2000eq,Nacher:2001yr,Mokeev:2001qg,Roca:2004vs,Fix:2005if,Roberts:2004mn,Kamano:2009im}, time is ripe to investigate  $\pi^0 \eta$ photoproduction,  which has brought us some surprises. 

$\pi^0 \eta$ photoproduction has been studied experimentally, and unpolarized and differential cross sections are reported in  Refs.~\cite{naka,ajaka,Horn:2007pp,Horn:2008qv,Kashevarov:2009ww}. Polarization observables are reported in Refs.~\cite{ajaka,Gutz:2008zz,Gutz:2009zh,Kashevarov:2010gk}. The reaction was studied theoretically in Refs.~\cite{Jido:2001nt,Doring:2005bx}. The result of Ref.~\cite{Doring:2005bx} was actually a prediction, though preliminary experimental results  from \cite{mariana} were available at that time. Following this work, other more recent papers have tackled the problem theoretically based on  models \cite{Fix:2008pv} or partial wave analysis of the  data \cite{Anisovich:2004zz,Anisovich:2007bq}.  In the work of~\cite{Doring:2005bx}  the process turned out to be dominated at low energies by the excitation of the $\Delta(1700)$, which then decays into  $\eta\Delta $, with the $\Delta$ subsequently decaying into $\pi N$.  Predictions of the cross section were made possible because the  $\Delta(1700)$ happens to be one of the dynamically generated resonances  from the interaction of pseudoscalar mesons with the baryons of the  $\Delta$ decuplet \cite{Kolomeitsev:2003kt,Sarkar:2004jh}.  The couplings  of the resonance to the different channels were calculated in \cite{Sarkar:2004jh} and these, together with the experimental know\-ledge  of the decay of the $\Delta(1700)$ into $\gamma N$, allowed one to obtain  absolute numbers for the cross section which agree with present measurements. 

In Ref.~\cite{Doring:2007rz} the radiative decay width of the $\Delta(1700) \to\gamma N$ could be predicted, because the photon coupling to the mesons and  baryons that constitute this resonance are all well known. The result is in  agreement with the phenomenologically known values~\cite{Amsler:2008zzb} from  data analyses which are used in Refs.~\cite{ajaka,Doring:2005bx,Doring:2006pt}  for the $\gamma p\to\pi^0\eta p$ reaction. In Ref.~\cite{Doring:2007rz}, also the $\pi N$ channel, that couples weakly to the $\Delta(1700)$ in $D$-wave,  has been included in the unitary model, and a good fit to the $\pi N$ phase  shifts could be obtained.

The theoretical framework from Refs.~\cite{Doring:2005bx,Sarkar:2004jh} is quite predictive  since another one of the couplings of the $\Delta(1700)$ resonance is to the  $K \Sigma(1385)$ state and one can evaluate the total and differential cross  sections for the reaction $\gamma p \to K^0 \pi^0 \Sigma^+$ \cite{Doring:2005bx}, which agree with the measurements published in 
Ref.~\cite{Nanova:2008kr}. 

\begin{figure}
\bc
\includegraphics[width=0.35\textwidth]{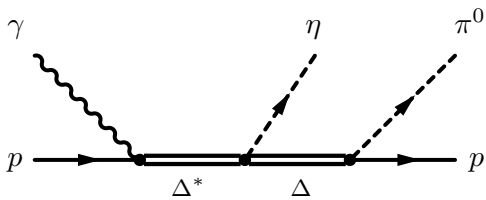}
\ec
\caption{Tree level process from the decay of the $\Delta(1700)$ to 
$\eta\Delta(1232)$. This is the dominant process (for the full list of 
processes, see Ref.~\cite{Doring:2005bx}). The complex 
$\Delta(1700)\to\eta\Delta(1232)$ coupling is a prediction within 
the chiral unitary framework.}
\label{fig:tree}
\end{figure}

\begin{figure}
\bc
\includegraphics[width=0.35\textwidth]{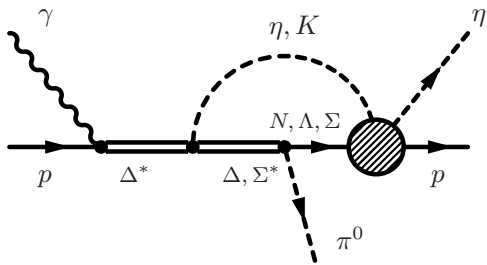}
\ec
\caption{Processes with $\Delta(1700)\eta\Delta$ and
  $\Delta(1700)K\Sigma(1385)$ 
couplings and $\pi^0\,S_{11}(\eta p)$ final states. These processes 
give a major contribution to the $\pi^0 N(1535)$ final state.}
\label{fig:otherchiral}
\end{figure}

From the list of the underlying processes of the theoretical framework  of Ref.~\cite{Doring:2005bx}, we show here only those of Figs.~\ref{fig:tree}  and \ref{fig:otherchiral}, which
involve the $\Delta(1700)\eta \Delta$ and  $\Delta(1700)K \\ \Sigma(1385)$ vertices predicted from the chiral unitary  amplitudes~\cite{Sarkar:2004jh}. These processes give the largest
contributions to the $\eta\Delta$ and $\pi^0 S_{11}(\eta p)$ final states.  The latter appears from the unitarization of the meson-baryon amplitude  in which the $N(1535)$ appears
dynamically generated. There are, however, also  processes given by $S$-channel resonance exchange taken from the $\gamma N\to \pi\pi N$ Valencia model of Ref.~\cite{Roca:2004vs}.
Furthermore, there are  contributions from Kroll-Ruderman and meson pole terms, contributions from the  normal and anomalous magnetic moments of the baryons, and combinations of  those
processes. All free constants that appear in the model have been fixed  from other processes, thus the results of Ref.~\cite{Doring:2005bx}  can be regarded as predictions.

The model resulting from Figs.~\ref{fig:tree} and \ref{fig:otherchiral} is gauge invariant. This is so 
since the $\gamma N \Delta(1700)$ coupling is obtained from the experimental 
data through an expression which is manifestly gauge invariant [see Eqs.~(72,73) of Ref.~\cite{Nacher:2000eq}]. The dominance of the 
$\Delta(1700)$ excitation is also corroborated by recent analyses of the 
data~\cite{Fix:2008pv,Anisovich:2004zz,Anisovich:2007bq,Fix:2010bd}. The full model of Ref.~\cite{Doring:2005bx} contains other terms, apart 
from the largely dominant ones of Figs.~\ref{fig:tree}, \ref{fig:otherchiral}. Some of these minor terms considered in Ref.~\cite{Doring:2005bx} involve a 
photon coupling to a loop and a meson baryon scattering amplitude 
obtained using the chiral unitary approach (see for instance Fig.~7 of 
\cite{Doring:2005bx}). It was proved in Refs.~\cite{Borasoy:2005zg,Borasoy:2007ku} (see also \cite{Nacher:1999ni}) that due to the series 
of diagrams implicit in the Bethe Salpeter expansion of the scattering 
matrix, gauge invariance of the model requires that the photon is 
coupled to particles and vertices in the intermediate loops and not just 
only to the last loop.
 
   Although formally needed for the test of gauge invariance, these 
terms are generally small. They vanish in the limit of low photon 
momentum, and induce corrections of order $(p_\gamma/2M_p)^2$, which must 
be evaluated in each case if $p_\gamma$ is not small. In the work of \cite{Doring:2005bx} 
they were evaluated and found to induce corrections of the order of 5\% 
of the contribution of the term with the photon coupled to the last loop 
considered in Fig. 7 of \cite{Doring:2005bx}.
Since this latter term represents a small 
correction compared to the contribution of the diagrams of Figs. \ref{fig:tree} and \ref{fig:otherchiral} 
here, neglecting the gauge fixing terms of \cite{Borasoy:2005zg,Borasoy:2007ku,Nacher:1999ni} in the present 
case, corresponds to neglecting small corrections to small terms and which
were safely disregarded in Ref.~\cite{Doring:2005bx} and in the present work.

The resulting theoretical framework also allows to correlate up to eleven different cross sections induced by photons or pions and this work has been reported in Ref.~\cite{Doring:2006pt}. All these successful
predictions strongly support   the claim of the $\Delta(1700)$  being a dynamically generated resonance. Yet,  there are more challenges for this approach, and the polarization observables 
of the $\vec{\gamma} p \to \pi^0 \eta p$ reaction are some of them. The polarization asymmetry $\Sigma$ was measured in Ref.~\cite{ajaka} and in the  same work the approach of Ref.~\cite{Doring:2005bx}
was used to describe the data.  It was found there that the theoretical framework made the right predictions and that the  presence of the $\Delta(1700)$ was essential for this success. 

The recent work of Ref.~\cite{Gutz:2009zh} presents another challenge since  new observables are measured, i.e, the $I^S$ and $I^C$ polarizations as a  function of the $\phi^*$ angle
between the decay plane and the reaction  plane [cf. Fig.~\ref{fig:ebenen}]. 

The purpose of the present work is to evaluate these new polarization
observables, and also $d\Sigma/d\cos\theta$ measured
recently~\cite{Gutz:2008zz}, 
in order to test the theoretical framework of 
Ref.~\cite{Doring:2005bx} and hence the nature of the $\Delta(1700)$.  As we shall
see, these observables are well reproduced and one can see that in the 
absence of the $\Delta(1700)$ term in the amplitude the results grossly 
deviate from experiment, showing once again the essential role played 
by this resonance in the $\gamma p \to \pi^0 \eta p$ reaction and providing 
further support for the idea of this resonance as being dynamically generated. 

The paper is organized as follows: In sec.~\ref{sec:formalism} we present the
formalism, followed by the presentation of our results in sec.~\ref{sec:res}.
We end with a short summary.

%%%%%%%%%%%%%%%%%%%%%%%%%%%%%%%%%%%%%%%%%%%%%%%%%%%%%%%%%%%%%%%%%%%%%%%%%%%%%%%%%%%%%%%%%%%%%%%%%

\section{Formalism}
\label{sec:formalism}
\subsection{Reaction geometry}
\label{sec:geometry}
\begin{figure*}\sidecaption
\includegraphics[width=0.72\textwidth]{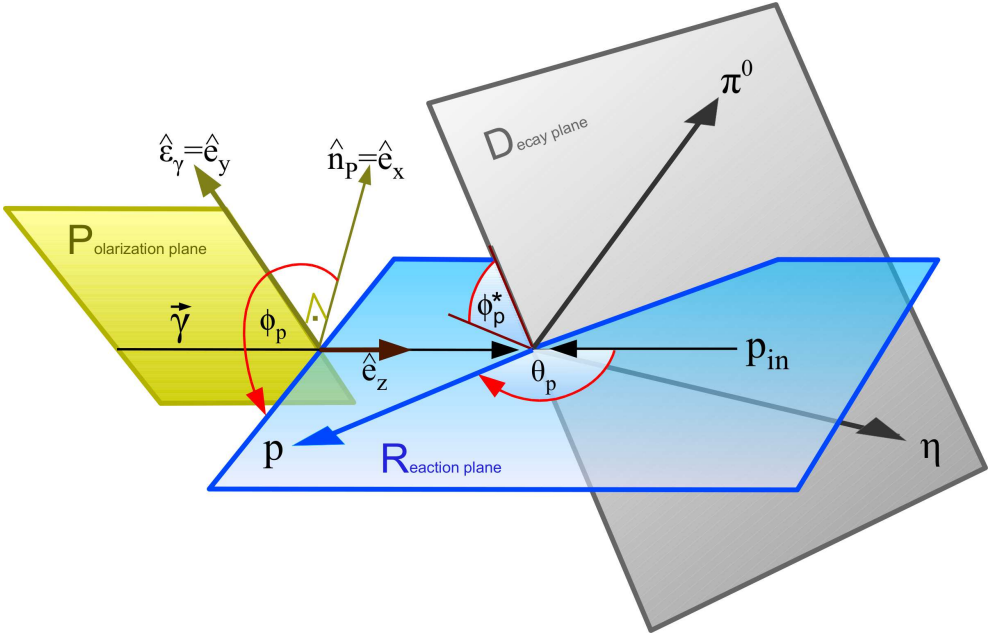}
\caption{Definition of the angles $\phi$ and $\phi^*$ (center-of-mass-frame). 
The polarization plane (P) is defined by the photon momentum $\vec{k}_\gamma$ 
and the polarization vector $\hat{\vec{\epsilon}}_\gamma$. Here, the photon is 
polarized in the vertical plane,
$\hat{\vec{\epsilon}}_\gamma||\hat{\vec{e}}_y$ 
and then the normal $\hat{\vec{n}}_P$ of the polarization plane is 
$\hat{\vec{n}}_P||\hat{\vec{e}}_x$. The reaction plane (R) is defined by 
$\vec{k}_\gamma$ and the momentum of the spectator, here the recoiling (final
state) proton with $\vec{p}_p$. The decay plane (D) is defined by the momenta 
of the final states, given by $\vec{p}_p$ and the $\pi^0$ and $\eta$ momenta 
$\vec{p}_{\pi^0}$ and $\vec{p}_\eta$. The polar angle of the
spectator/recoiling particle is denoted as $\theta_p$. The initial proton 
is given by $\vec{p}_{in}$.}
\label{fig:ebenen}
\end{figure*}

Fig.~\ref{fig:ebenen} shows the geometry of the reaction $\vec{\gamma}p
\to\pi^0\eta p$. All quantities are defined in the overall center-of-mass 
(c.m.) system. In the figure, the configuration is chosen in which the 
proton is the spectator; the two other cases of the $\pi^0$ and $\eta$ being
the spectator are defined analogously. The definitions given here agree with 
those of the CBELSA/TAPS experiment~\cite{Gutz:2009zh,gutzprivate}.

The photon momentum is in the $z$ direction, $\vec{k}_\gamma||\hat{\vec{e}}_z$. 
Its linear polarization is chosen in the $y$ direction, i.e. the polarization 
vector is $\hat{\vec{\epsilon}}_\gamma||\hat{\vec{e}}_y$. This geometry is 
usually referred to as polarization in the ``vertical plane''. Then, for the 
normal $\hat{\vec{n}}_P$ of the polarization plane P, $\hat{\vec{n}}_P||\hat{\vec{e}}_x$. 

The reaction plane R is defined by the momentum of the incoming photon 
$\vec{k}_\gamma$ and the momentum of the spectator, here the recoiling (final
state) proton with momentum $\vec{p}_p$. The angle $\phi$ is defined as the 
angle between the reaction plane and the normal to the polarization plane. 
This azimuthal angle is counted in the $xy$ plane counterclockwise starting 
from the direction $\hat{\vec{e}}_x$. For example, for a spectator proton of 
three momentum $\vec{p}_p=(0,y,z)$ ($y,z$ arbitrary), $\phi_p=+90^0$. The 
$\phi$ angles for the cases of $\pi^0$ and $\eta$ spectators are denoted as 
$\phi_{\pi^0}$ and $\phi_{\eta}$ in the following. The normal of the 
reaction plane R is given by
\be
\hat{\vec{n}}_R^p	
=\frac{\hat{\vec{e}}_z\times\vec{p}_p}{|\hat{\vec{e}}_z\times\vec{p}_p|},\,
\hat{\vec{n}}_R^{\pi^0} 
=\frac{\hat{\vec{e}}_z\times\vec{p}_{\pi^0}}{|\hat{\vec{e}}_z\times\vec{p}_{\pi^0}|},
\hat{\vec{n}}_R^\eta	
=\frac{\hat{\vec{e}}_z\times\vec{p}_\eta}{|\hat{\vec{e}}_z\times\vec{p}_\eta|},
\ee
where the index of $\hat{\vec{n}}_R$ indicates the spectator particle.

The decay plane D is defined by the momenta of the three final state 
particles $p,\,\pi^0, \,\eta$ which all lie in one plane due to momentum 
conservation. The normal $\hat{\vec{n}}_D$ of the decay plane D defines 
the orientation,
\be
\hat{\vec{n}}_D^p	
=\frac{\vec{p}_{\pi^0}\times\vec{p}_\eta}{|\vec{p}_{\pi^0}\times\vec{p}_\eta|},
\hat{\vec{n}}_D^{\pi^0} 
=\frac{\vec{p}_{p    }\times\vec{p}_\eta}{|\vec{p}_{p    }\times\vec{p}_\eta|},
\hat{\vec{n}}_D^\eta	
=\frac{\vec{p}_{\pi^0}\times\vec{p}_p   }{|\vec{p}_{\pi^0}\times\vec{p}_p   |}.
\ee
$\phi^*$ is the angle between the normals of the R and the D planes,
unambiguously defined as 
\be
\phi^*_i=
\begin{cases}
\hspace*{0.83cm} \arccos(\hat{\vec{n}}_D^i\cdot\hat{\vec{n}}_R^i) 	
& \text{if}\,\,\,(\hat{\vec{n}}_D^i\times\hat{\vec{n}}_R^i)\cdot\vec{p}_i>0\\
2\pi-\arccos(\hat{\vec{n}}_D^i\cdot\hat{\vec{n}}_R^i) 	
& \text{otherwise}
\end{cases}
\ee
for all three cases $i=p,\,\pi^0,\,\eta$. Note that $\hat{\vec{n}}_D^p\times
\hat{\vec{n}}_R^p$ and $\vec{p}_p$ are parallel, because $\vec{p}_p$ lies in 
both the R and the D planes (analogously for the $\pi^0$ and $\eta$ spectator 
cases). As we will also evaluate the beam asymmetry $d\Sigma/d\cos\theta$ 
measured in Ref.~\cite{Gutz:2008zz}, the polar angle $\theta_p$ of the
spectator  is also shown in Fig.~\ref{fig:ebenen}.

%%%%%%%%%%%%%%%%%%%%%%%%%%%%%%%%%%%%%%%%%%%%%%%%%%%%%%%%%%%%%%%%%%%%%%%%%%%%%%%%%%%%%%%%%%%%%%%%%

\subsection{Monte Carlo evaluation}
In this section, the details of the Monte Carlo evaluation of the three-body 
phase space are presented. With this method, different observables can be 
easily evaluated which would otherwise require a tedious re-parameteriza\-tion 
of the amplitude for every new measured observable. The total energy is $\sqrt{s}\equiv W$.

The total cross section of the $\gamma p\to\pi^0\eta p$ reaction in 
Ref.~\cite{Doring:2005bx}, Eq. (47), is written as an integral over the
$\pi^0p$ 
invariant mass, with one of the phase space integrals evaluated in 
the $\pi^0 p$ rest frame, 
\be
\sigma=\int\limits_{m_{\pi^0}+M_p}^{\sqrt{s}-m_\eta}\,
dM_I(\pi^0 p)\,\frac{d\sigma}{dM_I(\pi^0 p)}
\ee
with
\be
\frac{d\sigma}{dM_I(\pi^0 p)}&=&\frac{1}{4(2\pi)^5}\frac{M_p
M_f}{s-M_p^2}\;\frac{\tilde{p}_\pi
p_\eta}{\sqrt{s}}\int\limits_0^{2\pi}d\phi_\eta\int\limits_{-1}^1d\cos\theta_\eta\non
&\times&\int\limits_0^{2\pi}d\tilde{\phi}\int\limits_{-1}^1d\cos\tilde{\theta}\;
\overline{\sum}\sum |T_{\gamma p\to\eta\pi^0 p}|^2\non
\label{dsdmpidel}
\ee  
with $M_f=M_p$ and $\tilde{p}_\pi$ the modulus of the momentum  
${\vec {\tilde  p}}_\pi$ of the $\pi^0$ in the $\pi^0 p$ rest-frame, 
\be
{\tilde p}_\pi=\frac{1}{2M_I}\lambda^{1/2}(M_I^2,
m_\pi^2,M_p^2)
\ee
where $\lambda^{1/2}(a^2, b^2, c^2)=\sqrt{[a^2-(b+c)^2][a^2-(b-c)^2]}$ is the 
K\"all\'en function and the direction of ${\vec {\tilde p}}_\pi$  is given by 
$\tilde{\phi}$ and $\tilde{\theta}$. This vector is connected to  
${\vec p_\pi}$ in the $\gamma p$ rest-frame by the boost
\be  {\vec 
p}_\pi=\left[\left(\frac{\sqrt{s}-\omega_\eta}{M_I}-1\right)\left(-\frac{{\vec
{\tilde p}}_\pi{\vec p}_\eta}{{\vec p}_\eta^{\;2}}\right)+\frac{{\tilde
p}_\pi^0}{M_I}\right]\left(-{\vec p}_\eta\right)+{\vec {\tilde p}}_\pi\non
\label{boost2}
\ee  
where ${\tilde p}_\pi^0=\sqrt{{\vec {\tilde  p}}_\pi^{\;2}+m_\pi^2}$. The
$\eta$ three-momentum in Eq.~(\ref{dsdmpidel}) is in the overall c.m. frame 
of $\gamma p$ and is given by $p_\eta=\lambda^{1/2}(s,M_I^2,
m_\eta^2)/(2\sqrt{s})$ and the two angles $\phi_\eta,\theta_\eta$. The 
$\overline{\sum}\sum$ denotes the usual average and sum over the initial
and the final states, respectively. 

The parameterization of the cross section in this way allows for the
calculation of the $\pi^0 p$ invariant mass distribution, but not, e.g., 
for the other 2 cases of invariant masses, which would require a 
re-parameterization of the amplitude and phase space integrals. 

However, the integral can be solved by Monte Carlo integration which 
allows for a relatively easy evaluation of other observables. For this, 
it is advantageous to rewrite the integration over $dM_I$ in terms 
of $dp_\eta$ by means of
\be
M_I^2(\pi^0p)=s+m_\eta^2-2\sqrt{s}\sqrt{m_\eta^2+p_\eta^2}.
\ee
Then, evenly distributed random events are generated for all five integration 
variables $p_\eta,\, \phi_\eta, \, \theta_\eta,\, \tilde\phi$, and
$\tilde\theta$ within their limits given in Eq.~(\ref{dsdmpidel}) 
(the limits of the $p_\eta$ integration are $p_\eta\in [0,\,\lambda^{1/2}
(s,(m_\pi+M_p)^2,m_\eta^2)/(2\sqrt{s})]$). 

A random event $(j)$ is given by a set of these five values 
$(p^{(j)}_\eta,\, \phi^{(j)}_\eta, \, \theta^{(j)}_\eta,\, \tilde\phi^{(j)},
\tilde\theta^{(j)})$ and the corresponding contribution to the phase space integral, 
\be
R^{(j)}&=&\left(\frac{\sqrt{s}}{M_I(\pi^0p)}\;\frac{p^{(j)}_\eta}{\omega_\eta}\right)
\;\frac{1}{4(2\pi)^5}\frac{M_p
M_f}{s-M_p^2}\;\frac{\tilde{p}_\pi
p^{(j)}_\eta}{\sqrt{s}}\non &\times&
\overline{\sum}\sum |T_{\gamma p\to\eta\pi^0 p}|^2(M_I(\pi^0 p),
\theta^{(j)}_\eta,\phi^{(j)}_\eta,\tilde{\theta}^{(j)},\tilde{\phi}^{(j)})\non
\label{rj}
\ee
where the first term in parenthesis comes from the change of integration from 
$M_I(\pi^0 p)$ to $p_\eta$. The quantities $\omega_\eta$, $M_I(\pi^0 p)$, and 
$\tilde{p}_\pi$ can all be expressed in terms of the five integration 
variables, as previously given. 

The total cross section is evaluated with a Monte Carlo integral that is given 
by the sum over events $R^{(j)}$, multiplied with the integration ranges for 
each variable, and divided by the number of events $N$,
\be
\sigma&=&\frac{1}{N}\;\sum_{j=1}^{N} E^{(j)},\non
E^{(j)}&=&(2\pi)^2\,2^2\,\frac{1}{2\sqrt{s}}\;
\lambda^{1/2}(s,m_\pi+M_p,m_\eta)\;R^{(j)}.
\label{ej}
\ee
Note that for the present study, we need an event set of polarized photons 
instead of unpolarized ones. The reaction amplitude can be written explicitly 
as $T_i=\hat{\vec{\epsilon}}_i \cdot\vec{T}$ where $i$ stands for the
polarization vectors in $x$ and $y$ direction. To get the total unpolarized 
cross section, one sums over $i$; thus, one obtains events polarized in the
$y$ direction by restricting the sum to the term $T_{i=y}$. Note that a factor 
of two has to be supplied to the cross section for polarized photons 
(compared to the unpolarized case) from the average over initial 
states $\overline{\sum}$ of Eq.~(\ref{dsdmpidel}).

With the method described above, it is now easy to evaluate many observables, 
because the only remaining task is to bin the random events $R^{(j)}$ to the 
bins of the desired observable, and those bins can be uniquely constructed 
from $(p^{(j)}_\eta,\, \phi^{(j)}_\eta, \, \theta^{(j)}_\eta,\,
\tilde\phi^{(j)},\tilde\theta^{(j)})$ for each event. 

For example, for the calculation of $I^S$ and $I^C$,  $N=4.8\cdot 10^5$ random
events are binned in $16\times 16$ bins for $\phi$ and $\phi^*$. For this, 
the three-vectors of the $p$, $\pi^0$, and $\eta$ in the overall c.m. system 
are reconstructed for each event from the five variables and using
Eq.~(\ref{boost2}). Then, $\phi^{(j)}$ and $(\phi^*)^{(j)}$ are calculated 
from the equations given in Sec.~\ref{sec:geometry} for each event. 

Finally, the binned double differential cross section for the bins 
$\phi_{i}$ and $\phi^*_k$ is given by
\be
\frac{d^2\sigma(\phi,\phi^*)}{d\phi\,d\phi^*}\simeq  
\frac{\Delta^2\sigma(\phi_{i},\phi^*_k)}{\Delta\phi\,\Delta\phi^*}
=\sum_{j=1}^N \frac{E^{(j)}\,f(i,j,k)}{\Delta\phi\,\Delta\phi^*\,N}
\label{binned}
\ee
with the weight $E^{(j)}$ from Eq.~(\ref{ej}), the bin widths $\Delta\phi$ and 
$\Delta\phi^*$, and $f(i,j,k)$ a function which is one if the event angles 
$\phi^{(j)}$ and $(\phi^*)^{(j)}$ are simultaneously in the bins $\phi_{i}$
and $\phi^*_k$, and zero otherwise. In practice, for each event of the Monte
Carlo run we determine the range (denoted the box) of $\phi_i$ and $\phi^*_k$ 
where the event is generated and accumulate the values 
$E^{(j)}$ in the respective boxes, such that in one run one obtains the 
double differential cross section.

For fixed $\phi^*$ and a fully linearly polarized beam, the $\phi$
differential cross section obeys the $\phi$ dependence~\cite{Roberts:2004mn,Gutz:2009zh}
\be
\frac{d\sigma}{d\phi}=\sigma_0 [1+I^S \sin(2\phi)+I^C \cos(2\phi)].
\label{phidep}
\ee
The last step to obtain $I^S$ and $I^C$ as a function of $\phi^*$ is to fit
the differential cross section of Eq.~(\ref{binned}) with the $\phi$
distribution given in Eq.~(\ref{phidep}), separately for every $\phi^*$ bin. 

One has to choose the binning slightly differently to evaluate another 
observable, measured in Ref.~\cite{Gutz:2008zz} and given by
\be
I^\theta=\frac{d\Sigma}{d\,\cos\theta}
\label{Itheta}
\ee
with $\theta$ being the polar angle of the spectator particle (cf. 
Fig.~\ref{fig:ebenen}). For this, we construct a double differential 
cross section in analogy to Eq. (\ref{binned}),
\be
\frac{d^2\sigma(\phi,\cos\theta)}{d\phi\,d\cos\theta}\simeq  
\frac{\Delta^2\sigma(\phi_{i},(\cos\theta)_k)}{\Delta\phi\,\Delta\cos\theta}
\ee
and fit it for fixed $\cos\theta$ with the ansatz
\be
\frac{d\sigma}{d\phi}=\sigma_0 [1+I^{S,\theta} \sin(2\phi)+I^\theta \cos(2\phi)]
\label{Itheta2}
\ee
and for the three cases of the $p$, $\pi^0$, and $\eta$ being the spectator.

The method described in this section has been tested extensively: Generating 
events with unpolarized photons: $I^S$ and $I^C$ are zero for all $\phi^*$, 
as must be. Also, we have checked the results of the Monte Carlo method 
with the $\eta p$ invariant mass distribution and the total cross section, 
that can be evaluated without the Monte Carlo method (cf. Eq.~(45) of 
Ref.~\cite{Doring:2005bx}). In fact, in the same binning algorithm used 
to obtain $I^S$ and $I^C$, the previously evaluated results from 
Refs~\cite{Doring:2005bx,ajaka} have been reproduced.  

%%%%%%%%%%%%%%%%%%%%%%%%%%%%%%%%%%%%%%%%%%%%%%%%%%%%%%%%%%%%%%%%%%%%%%%%%%%%%%%%%%%%%%%%%%%%%%%%%

\section{Results}
\label{sec:res}

\begin{figure*}\sidecaption
\includegraphics[width=0.7\textwidth]{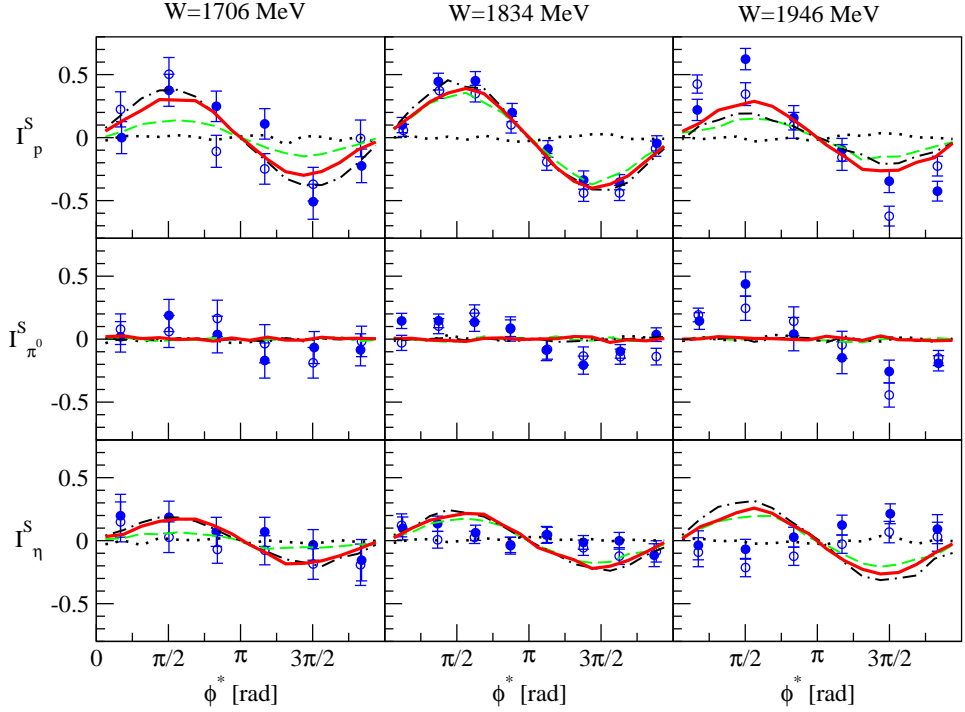}
\caption{(Color online) Polarization observable $I^S(\phi^*)$ for the three cases of $p$, 
$\pi^0$, and $\eta$ spectators and for three different energies
$W\equiv\sqrt{s}$. The data $I^S(\phi^*)$ are from Ref.~\cite{Gutz:2009zh} 
(full circles). The empty circles show $-I^S(2\pi-\phi^*)$. (Red) solid lines:
Present results, predicted from the model of
Refs.~\cite{Doring:2005bx,Doring:2006pt}. (Black) dotted lines: Without the 
$\Delta(1700)\eta\Delta$ and $\Delta(1700)K\Sigma(1385)$ couplings 
predicted from the chiral unitary model, i.e. without the processes from 
Figs.~\ref{fig:tree} and \ref{fig:otherchiral}. (Green) dashed lines: Without 
the contributions from Fig.~\ref{fig:otherchiral}. (Black) dash-dotted lines: 
Only contribution from Fig.~\ref{fig:tree}.}
\label{fig:is}
\end{figure*}

\begin{figure*}\sidecaption
\includegraphics[width=0.7\textwidth]{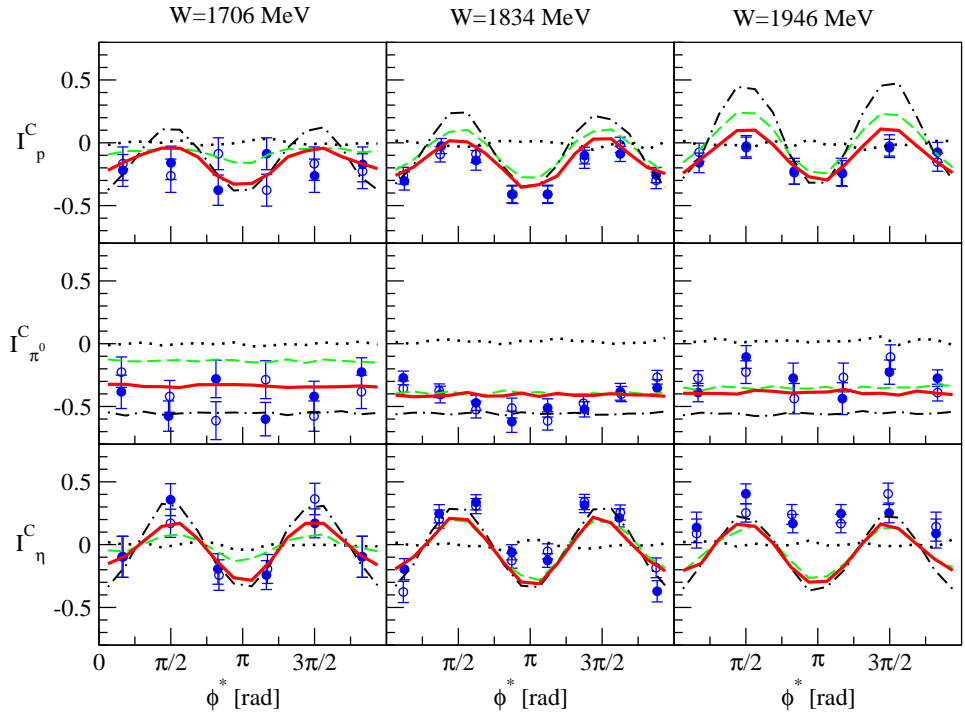}
\caption{(Color online) Polarization observable $I^C(\phi^*)$ for the three cases of $p$,
  $\pi^0$, and $\eta$ spectators. The data $I^C(\phi^*)$ are from 
Ref.~\cite{Gutz:2009zh}. The empty circles show $I^C(2\pi-\phi^*)$. 
(Red) solid lines: Present results, predicted from the model of 
Refs.~\cite{Doring:2005bx,Doring:2006pt}. 
The other curves are labeled as in Fig.~\ref{fig:is}.}
\label{fig:ic}
\end{figure*}

The measurements of $I^S$ and $I^C$ from Ref.~\cite{Gutz:2009zh} 
are shown in Figs.~\ref{fig:is} and \ref{fig:ic} (filled circles). 
These polarization observables obey the symmetries $I^S(\phi^*)
=-I^S(2\pi-\phi^*)$ and $I^C(\phi^*)=I^C(2\pi-\phi^*)$ \cite{Gutz:2009zh}. 
The experimental data according to these relations are shown with the open 
circles in Figs.~\ref{fig:is} and \ref{fig:ic}; the discrepancy to the 
original data set (filled symbols) is within the statistical error 
estimated in Ref.~\cite{Gutz:2009zh}.

The results of the present study for $I^S$ and $I^C$ are shown in 
Figs.~\ref{fig:is} and \ref{fig:ic} with the (red) solid lines~\footnote{Note 
that for the lowest energy $W=1706$~MeV, the theoretical total cross 
section is raising much faster than linear~\cite{ajaka}, and the model 
has not been evaluated at the center of the experimental bin at $W=1706$~MeV, 
but at $W=1740$~MeV which corrects for the finite experimental bin width. 
The value of $W=1740$~MeV has been obtained by calculating the expectation 
value of $k_\gamma$, using the theoretical total cross section in the 
experimental bin over the energy range of $1706\pm 64$~MeV.}, together 
with the data from Ref.~\cite{Gutz:2009zh}.
At a given total energy $W\equiv\sqrt{s}$, $4.8\cdot 10^5$ Monte Carlo events
have been binned in $16\times 16$ bins for $\phi$ and $\phi^*$, and the
quantities $I^S$ and $I^C$ have been obtained by fitting the $\phi$
distributions as described in the previous section. The binning is clearly 
visible in Figs.~\ref{fig:is} and \ref{fig:ic}, for which the theoretical
results for the 16 $\phi^*$ bins have been connected with piecewise straight lines.

The agreement with the data in Figs.~\ref{fig:is} and \ref{fig:ic} is good for 
the two lower energies $W=1706$~MeV and $W=1834$~MeV, given that the
theoretical curves are predictions of the model from 
Refs.~\cite{Doring:2005bx,Doring:2006pt}. For the highest energy $W=1946$~MeV, 
deviations of the prediction from the data start to become noticeable in 
$I^S_{\pi^0}$, $I^S_{\eta}$ and $I^C_{\eta}$, while $I^S_{p}$ and $I^C_{p}$
are still well predicted. This is a sign that at higher energies new
mechanisms start to become important which are not considered in 
Refs.~\cite{Doring:2005bx,Doring:2006pt}. This could, e.g., be a 
$\Delta(1940)\,D_{33}$ resonance which was needed in the Bonn-Gatchina partial 
wave analysis of the same reaction in Ref.~\cite{Horn:2007pp}. Also, the
$a_0(980) p$ final state, which is not included in the present model, can play 
a role at higher energies. It has been clearly seen at high energies in
invariant mass distributions of the reaction~\cite{gutzprivate}.

To test the sensitivity of different processes to $I^S$ and $I^C$, we have
removed parts of the theoretical model and re-evaluated  $I^S$ and $I^C$. From
the list of processes discussed in Ref.~\cite{Doring:2005bx}, the ones shown
in Figs.~\ref{fig:tree} and \ref{fig:otherchiral} are of special interest, 
because they contain the non-trivial predictions of the chiral unitary model 
of Ref.~\cite{Sarkar:2004jh} for the large, complex coupling constants of the 
$\Delta(1700)$ to $\eta\Delta$ and $K\Sigma(1385)$. Omitting these processes,
the (black) dotted lines in Figs.~\ref{fig:is} and \ref{fig:ic} are obtained, in gross 
disagreement with the data. This shows that these processes are indeed
responsible for the $\phi^*$ dependence and the good agreement with the data. 
Note also that for the dotted curves, small effects of statistical noise
become clearly visible, tied to the Monte Carlo based evaluation of the phase 
space integrals discussed in the previous section.

To test the sensitivity of the processes shown in Fig.~\ref{fig:otherchiral}, 
we have switched them off. The pertinent results are shown with the (green) dashed 
lines in Figs.~\ref{fig:is} and \ref{fig:ic}
and turn out to be close to the full solution. This suggests that the process 
from Fig.~\ref{fig:tree} is dominant. The result from this process alone is
shown with the (black) dash-dotted lines in Figs.~\ref{fig:is} and
\ref{fig:ic}. Indeed, the data are still qualitatively described, although the 
contribution from this process alone tends to overshoot the $\phi^*$
dependence of the result (cf., e.g., $I^C_p$ at $W=1834$~MeV and
$W=1946$~MeV). 
The latter finding implies that this process, although dominant, requires the 
presence of other processes with weaker $\phi^*$ dependence, such as given in 
Fig.~\ref{fig:otherchiral} and also by the rest of the contributions discussed 
in Ref.~\cite{Doring:2005bx} not displayed here but included. Then, the full
solution, that contains all these processes and their interferences, leads to 
the good prediction of the data.

\begin{figure}
\includegraphics[width=0.48\textwidth]{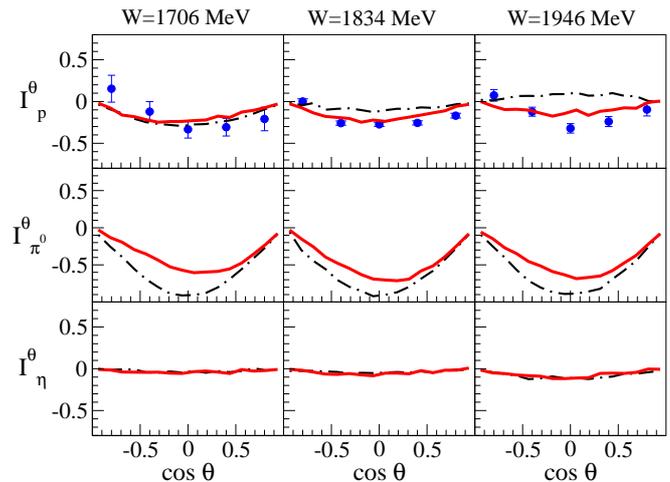}
\caption{(Color online) Polarization observable $\displaystyle{I^\theta(\cos\theta)}=
\frac{d\Sigma}{d\,\cos\theta}$ for the three cases of $p$, $\pi^0$, and $\eta$ 
spectators and for three different energies $W\equiv\sqrt{s}$. The data are
from Ref.~\cite{Gutz:2008zz}. (Red) solid lines: Present results predicted from the 
model of Refs.~\cite{Doring:2005bx,Doring:2006pt}. (Black) dash-dotted lines: 
Only contribution from Fig.~\ref{fig:tree}.}
\label{fig:itheta}
\end{figure}
For completeness, also the observable $I^\theta = d\Sigma / d\,\cos\theta$, 
recently measured~\cite{Gutz:2008zz} by the CBELSA/TAPS collaboration, is 
evaluated in this study. In Ref.~\cite{Gutz:2008zz}, also data on the
distributions $d\Sigma/d\,M_I$ were provided, in general agreement with 
the measurements by the GRAAL collaboration in Ref.~\cite{ajaka} where one 
can also find the theoretical results of the present model for $d\Sigma/d\,M_I$. 

The results of the present model for $I^\theta$ [cf. Eqs.~(\ref{Itheta},
\ref{Itheta2})] are shown in Fig.~\ref{fig:itheta} (red solid lines). Also, the distributions for the $\pi^0$ and $\eta$ spectator cases are shown, where no data
are provided in Ref.~\cite{Gutz:2008zz}. The distributions for these two cases 
have already been analyzed and will be published in the near future~\cite{gutzpreparation}. 

The model describes the data quite well, although at the highest energy, clear 
deviations become noticeable, which was also the case for $I^S$ and $I^C$
shown before. The contribution from the diagram of Fig.~\ref{fig:tree} alone
is shown by the (black) dash-dotted lines. The proton spectator case $I^\theta_p$ is
of special interest, because unlike in the case of $I^S$ and $I^C$, at the 
energy of $W=1834$~MeV the contribution from Fig.~\ref{fig:tree} is small and 
the result comes mainly from the other processes of the
model of Ref.~\cite{Doring:2005bx}. This clearly shows that there are processes 
beyond the one of Fig.~\ref{fig:tree}, and the binning in $\cos\theta$ is sensitive to them. 

The results for the polarization $I^{S,\theta}$, defined in
Eq.~(\ref{Itheta2}), are zero within the statistics provided by the Monte
Carlo evaluation. This has been also found in experiment~\cite{Gutz:2008zz}.

%%%%%%%%%%%%%%%%%%%%%%%%%%%%%%%%%%%%%%%%%%%%%%%%%%%%%%%%%%%%%%%%%%%%%%%%%%%%%%%%%%%%%%%%%%%%%%%%%%%%%%%%%%%%

\subsection{Systematic uncertainties}

It is of utmost importance to discuss the theoretical uncertainty of the
results discussed so far, because only then one can truly speak of agreement 
(or disagreement) with the data. This is usually the hardest part of the
theoretical calculation. For the case at hand, systematic theoretical errors 
can only be estimated, as we discuss in what follows. Most interesting is the 
dependence on the couplings of the $\Delta(1700)$ to the $\eta\Delta$ and 
$K\Sigma(1385)$ channels, $g_{\eta
  \Delta}=1.7-1.4\,i,\,g_{K\Sigma(1385)}=3.3+0.7\,i$, 
which are predictions of the chiral unitary model of Ref.~\cite{Sarkar:2004jh}. 

\begin{figure*}\sidecaption
\includegraphics[width=0.7\textwidth]{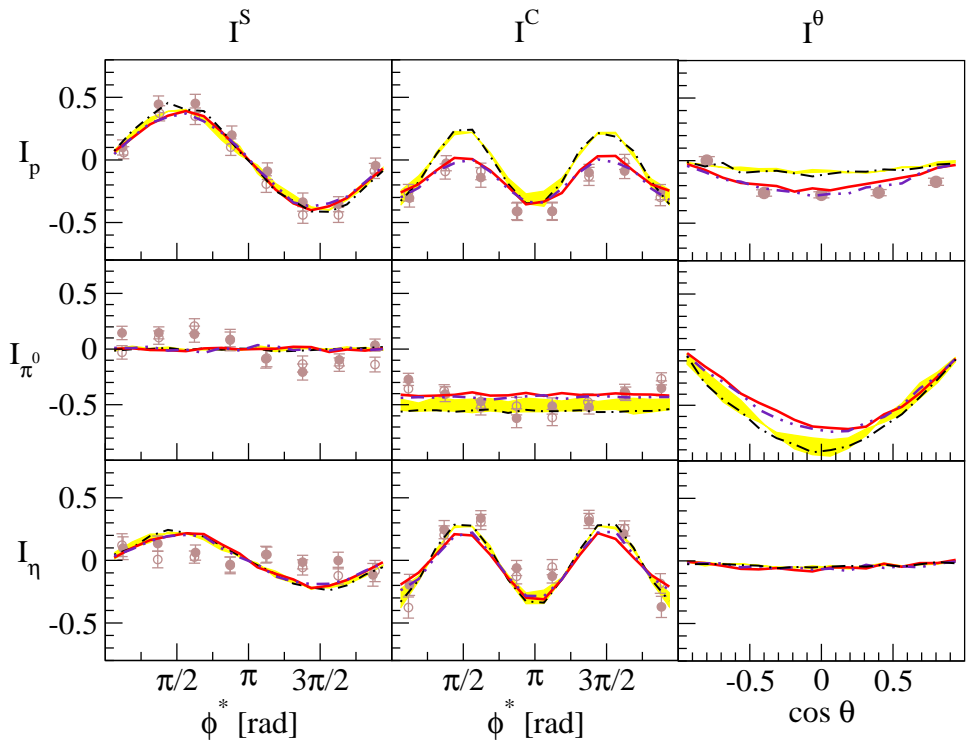}
\caption{(Color online) Estimate of the systematic error of the present model at 
$W=1834$~MeV. The (red) solid lines show the result of the full model of 
Ref.~\cite{Doring:2005bx}, the (indigo) dash-double dotted lines the full model, 
but with the updated $\Delta(1700)\eta\Delta$ and $\Delta(1700)K\Sigma(1385)$ 
coupling constants from Ref.~\cite{Doring:2007rz}. The (black) dash-dotted lines show 
the contribution from Fig.~\ref{fig:tree}. The (yellow) bands show the 
uncertainties of this contributions induced from the experimental 
uncertainties of the photon coupling of the $\Delta(1700)$.}
\label{fig:ierrors}
\end{figure*}

In Ref.~\cite{Doring:2007rz}, this model has been improved by the inclusion 
of the $\pi N$ channel in $D$-wave and a fit of the subtraction constants to 
the $\pi N$ $D_{33}$ phase shift of Ref.~\cite{Arndt:2006bf}. The updated
values of the coupling constants from Ref.~\cite{Doring:2007rz} are 
$g_{\eta \Delta}=-2.27-1.89\,i,\,g_{K\Sigma(1385)}=3.01+1.95\,i$. We have
evaluated the asymmetries $I^S$, $I^C$, and $I^\theta$ at the energy
$W=1834$~MeV with these updated values. The result is shown in
Fig.~\ref{fig:ierrors} with the (indigo) dash-double dotted lines.
Although the updated couplings $g$ are quite different from the original
values,  the asymmetries $I$ are almost unchanged: For $I^S$ and $I^C$, the
process of Fig.~\ref{fig:tree} is dominant, and the contributions to $I^S$ and
$I^C$ from this diagram alone are independent of the value of the coupling 
$g_{\eta \Delta}$ as the definition of Eq.~(\ref{phidep}) shows and taking
into account that $g_{\eta \Delta}$ factorizes in the amplitude of
Fig.~\ref{fig:tree}. Note, however, that $g_{\eta \Delta}$ and
$g_{K\Sigma(1385)}$  also appear in the rescattering diagrams shown in 
Fig.~\ref{fig:otherchiral}, but their influence on $I^S$ and $I^C$ is small.

Another source of uncertainty is the coupling of the photon to the 
$\Delta(1700)$, as shown in Fig.~\ref{fig:tree}. The $\gamma N\Delta(1700)$ 
phototransition is given by the helicity amplitudes $A_{1/2}=0.104\pm
0.015$~GeV$^{-1/2}$ 
and $A_{3/2}=0.085\pm 0.022$~GeV$^{-1/2}$ at the photon point 
$Q^2=0$~\cite{Amsler:2008zzb}. These values translate into the couplings 
$g_1'$ and $g_2'$ in the amplitude [cf. Eqs. (39,43) of
Ref.~\cite{Doring:2005bx}]. 
As in case of the strong couplings $g$ discussed above, for the process from 
Fig.~\ref{fig:tree} the asymmetries $I$ do not depend on the overall magnitude
of the $\gamma N\Delta(1700)$ coupling, but only on the ratio $g_1'/g_2'=-0.26 
M_N^{-1}/(+0.27 M_N^{-2})$ \cite{Nacher:2000eq,Doring:2005bx}. The mentioned 
uncertainties in $A_{1/2}$ and $A_{3/2}$ translate into an estimated
uncertainty of about $22\%$ for the ratio $g_1'/g_2'$. Evaluating the process
of Fig.~\ref{fig:tree} with the ratios $g_1'/g_2'(1\pm 0.22)$, we obtain the 
(yellow) bands shown in Fig.~\ref{fig:ierrors}. The result with the unchanged 
ratio $g_1'/g_2'$ is shown with the (black) dash-dotted lines. As the figures shows, 
the uncertainty from this source is well controlled.

Summarizing, we have shown that the $\Delta(1700)$ photoexcitation with
subsequent decay into $\eta\Delta$ as shown in Fig.~\ref{fig:tree} is the most 
important contribution to $I^S$ and $I^C$; however, we have also seen that the 
asymmetries $I^S$ and $I^C$ from this process are insensitive to the size
of the $\gamma N\Delta(1700)$ and $\Delta(1700)\eta\Delta$ couplings, and that 
systematic uncertainties in these couplings are well controlled.

%%%%%%%%%%%%%%%%%%%%%%%%%%%%%%%%%%%%%%%%%%%%%%%%%%%%%%%%%%%%%%%%%%%%%%%%%%%%%%%%%%%%%%%%%%%%%%%%%

\subsection{Interconnections with other experiments}

One should see the present results also in perspective to the 
previously evaluated observables from
Refs.~\cite{Doring:2005bx,Doring:2006pt,ajaka}. 
In the first work of Ref.~\cite{Doring:2005bx}, where the theoretical
framework was developed, the total cross section and invariant masses 
were predicted. While at the lowest energies the theoretical cross section 
is slightly below the data~\cite{ajaka}, the overall agreement is good, 
in particular also for the case of the related reaction $\gamma p \to
\pi^0 K^0\Sigma^+$: the predictions for the differential and total cross 
sections have been experimentally confirmed recently in Ref.~\cite{Nanova:2008kr}. 

However, for some of the $\eta p$ invariant masses in $\gamma p\to\pi^0\eta
p$, deviations from the data have been observed in Ref.~\cite{ajaka}, which 
are most probably tied to a too narrow $N(1535)$ which also appears
dynamically generated in the present model, based on the study of 
Ref.~\cite{Inoue:2001ip}. This problem appears also in the different 
context of single pion photoproduction and has been recently addressed in 
Refs.~\cite{Doring:2009uc,Doring:2009qr}. It was traced back to the absence 
of the $N(1650)\,S_{11}$ resonance in the original model~\cite{Inoue:2001ip} 
which was then included in Ref.~\cite{Doring:2009uc} and led to a consistent 
description of pion- and photon-induced single pion production. An upgrade of 
the present model for $\gamma p\to\pi^0\eta p$ with these new results would be desirable. 

For the other invariant masses in $\gamma p\to\pi^0\eta p$, in particular 
$M_I(\pi^0\eta)$, better agreement with the GRAAL experiment has been observed 
in Ref.~\cite{ajaka}. There, also the predictions for the beam asymmetries 
$d\Sigma/dM_I$ have been compared to experiment and an overall agreement could 
be found. Note that the beam asymmetries $\Sigma$, which have also been
measured in the CBELSA/TAPS experiment~\cite{Gutz:2008zz}, correspond to 
$I^C$ if the dependence on $\phi^*$ is integrated out, as has been pointed 
out in Refs.~\cite{Gutz:2009zh,gutzprivate}.

The formalism developed in Ref.~\cite{Doring:2005bx} has also been applied to 
other reactions in Ref.~\cite{Doring:2006pt}. In particular, the
predicted~\cite{Sarkar:2004jh} 
$\Delta(1700)$ couplings to $\eta\Delta$, $K\Sigma(1385)$, and $\pi\Delta$ 
could be used to evaluate eleven different pion- and photon-induced reactions 
resulting in an overall agreement, although for some of the reactions studied, 
some extra ingredients would be needed at higher energies, where a strong
forward peaking of some differential cross sections becomes noticeable.

%%%%%%%%%%%%%%%%%%%%%%%%%%%%%%%%%%%%%%%%%%%%%%%%%%%%%%%%%%%%%%%%%%%%%%%%%%%%%%%%%%%%%%%%%%%%%%%%%

\section{Conclusions}
The theoretical predictions of $I^S$, $I^C$, and $I^\theta$ in the $\gamma
p\to\pi^0\eta p$ reaction, evaluated from a chiral unitary model, agree well 
with the data recently measured at CBELSA/TAPS. We have shown that from the 
list of processes discussed in Ref.~\cite{Doring:2005bx}, the photon
excitation of the $\Delta(1700)$ with subsequent $S$-wave decay into 
$\eta\Delta$ is mainly responsible for the $\phi^*$ dependence. 
The $\Delta(1700)$ couplings to $\eta\Delta$, $K\Sigma(1385)$, and $\pi\Delta$ 
are predictions from a chiral unitary model in which the $\Delta(1700)$
appears dynamically generated from the unitarized interaction of the octet 
of $J^P=1/2^-$ mesons with the decuplet of $J^P=3/2^+$ baryons. The present 
results provide further evidence for this concept, which had been previously 
tested and verified for other observables in $\gamma p\to\pi^0\eta p$ and 
also in ten other pion- and photon-induced reactions.

%%%%%%%%%%%%%%%%%%%%%%%%%%%%%%%%%%%%%%%%%%%%%%%%%%%%%%%%%%%%%%%%%%%%%%%%%%%%%%%%%%%%%%%%%%%%%%%%%

\begin{acknowledgement}
This work is partly supported by DGICYT and FEDER funds Contract No. FIS 2006-03438, the
Generalitat Valenciana in the program Prometeo and the EU Integrated 
Infrastructure Initiative Hadron Physics Project under contract
RII3-CT-2004-506078. This work is also supported by the DFG (Deut\-sche
Forschungsgemeinschaft, Gz: DO 1302/1-2 and SFB/TR-16). We would like to 
thank E.~Gutz for providing the information on the definitions of the 
reaction geometry, useful discussions, and a careful reading of the manuscript.
\end{acknowledgement}


\begin{thebibliography}{99}
\bibitem{Braghieri:1994rf}
  A.~Braghieri {\it et al.},
  %``Total Cross-Section Measurement For The Three Double Pion Production
  %Channels On The Proton,''
  Phys.\ Lett.\  B {\bf 363}, 46 (1995).
  %%CITATION = PHLTA,B363,46;%%
\bibitem{Zabrodin:1997xd}
  A.~Zabrodin {\it et al.},
  %``Total cross section measurement of the gamma n --> p pi- pi0 reaction,''
  Phys.\ Rev.\  C {\bf 55}, 1617 (1997).
  %%CITATION = PHRVA,C55,1617;%%
\bibitem{Harter:1997jq}
  F.~H\"arter {\it et al.},
  %``Two neutral pion photoproduction off the proton between threshold and
  %800-MeV,''
  Phys.\ Lett.\  B {\bf 401}, 229 (1997).
\bibitem{Zabrodin:1999sq}
  A.~Zabrodin {\it et al.},
  %``Invariant mass distributions of the gamma n --->p pi- pi0 reaction,''
  Phys.\ Rev.\  C {\bf 60}, 055201 (1999).
  %%CITATION = PHRVA,C60,055201;%%
  %%CITATION = PHLTA,B401,229;%%
\bibitem{Wolf:2000qt}
  M.~Wolf {\it et al.},
  %``Photoproduction of neutral pion pairs from the proton,''
  Eur.\ Phys.\ J.\  A {\bf 9}, 5 (2000).
  %%CITATION = EPHJA,A9,5;%%
\bibitem{Langgartner:2001sg}
  W.~Langgartner {\it et al.},
  %``Direct Observation Of A Rho Decay Of The D(13)(1520) Baryon Resonance,''
  Phys.\ Rev.\ Lett.\  {\bf 87}, 052001 (2001).
  %%CITATION = PRLTA,87,052001;%%
\bibitem{Assafiri:2003mv}
  Y.~Assafiri {\it et al.},
  %``Double Pi0 Photoproduction On The Proton At Graal,''
  Phys.\ Rev.\ Lett.\  {\bf 90}, 222001 (2003).
  %%CITATION = PRLTA,90,222001;%%
\bibitem{Ripani:2002ss}
  M.~Ripani {\it et al.}  [CLAS Collaboration],
  %``Measurement of e p --> e' p pi+ pi- and baryon resonance analysis,''
  Phys.\ Rev.\ Lett.\  {\bf 91}, 022002 (2003).
  %[arXiv:hep-ex/0210054].
  %%CITATION = PRLTA,91,022002;%%
\bibitem{Ahrens:2003na}
  J.~Ahrens {\it et al.}  [GDH and A2 Collaborations],
  %``Helicity dependence of the gamma(pol.) p(pol.) $\to$ n pi+ pi0 reaction in
  %the second resonance region,''
  Phys.\ Lett.\  B {\bf 551}, 49 (2003).
  %%CITATION = PHLTA,B551,49;%%
\bibitem{Kotulla:2003cx}
  M.~Kotulla {\it et al.},
  %``Double pi0 photoproduction off the proton at threshold,''
  Phys.\ Lett.\  B {\bf 578}, 63 (2004).
  %[arXiv:nucl-ex/0310031].
  %%CITATION = PHLTA,B578,63;%%
\bibitem{Ahrens:2005ia}
  J.~Ahrens {\it et al.}  [GDH and A2 Collaborations],
  %``Intermediate Resonance Excitation In The Gamma P $\to$ P Pi0 Pi0
  %Reaction,''
  Phys.\ Lett.\  B {\bf 624}, 173 (2005).
  %%CITATION = PHLTA,B624,173;%%
\bibitem{Strauch:2005cs}
  S.~Strauch {\it et al.}  [CLAS Collaboration],
  %``Beam-helicity asymmetries in double charged pion photoproduction on the
  %proton,''
  Phys.\ Rev.\ Lett.\  {\bf 95}, 162003 (2005).
  %[arXiv:hep-ex/0508002].
  %%CITATION = PRLTA,95,162003;%%
\bibitem{Ahrens:2007zzj}
  J.~Ahrens {\it et al.}  [GDH and A2 Collaborations],
  %``First measurement of the helicity dependence for the gamma p $\to$ p pi+
  %pi- reaction,''
  Eur.\ Phys.\ J.\  A {\bf 34}, 11 (2007).
  %%CITATION = EPHJA,A34,11;%%
\bibitem{Thoma:2007bm}
  U.~Thoma {\it et al.},
  %``N* and Delta* decays into N pi0 pi0,''
  Phys.\ Lett.\  B {\bf 659}, 87 (2008).
  %[arXiv:0707.3592 [hep-ph]].
  %%CITATION = PHLTA,B659,87;%%
\bibitem{:2007bk}
  A.~V.~Sarantsev {\it et al.},
  %``New results on the Roper resonance and the $P_{11}$ partial wave,''
  Phys.\ Lett.\  B {\bf 659}, 94 (2008).
  %[arXiv:0707.3591 [hep-ph]].
  %%CITATION = PHLTA,B659,94;%%
\bibitem{:2009fq}
  M.~Battaglieri {\it et al.}  [CLAS Collaboration],
  %``Photoproduction of $\pi^+ \pi^-$ meson pairs on the proton,''
  Phys.\ Rev.\  D {\bf 80}, 072005 (2009).
  %%CITATION = PHRVA,D80,072005;%%
\bibitem{Krambrich:2009te}
  D.~Krambrich {\it et al.}  [Crystal Ball at MAMI Collaboration, TAPS
           Collaboration and A2 Collaboration],
  %``Beam-Helicity Asymmetries in Double Pion Photoproduction off the Proton,''
  Phys.\ Rev.\ Lett.\  {\bf 103}, 052002 (2009).
  %[arXiv:0907.0358 [nucl-ex]].
  %%CITATION = PRLTA,103,052002;%%
\bibitem{GomezTejedor:1993bq}
  J.~A.~G\'omez Tejedor and E.~Oset,
  %``A Model For The Gamma P $\to$ Pi+ Pi- P Reaction,''
  Nucl.\ Phys.\  A {\bf 571}, 667 (1994).
  %%CITATION = NUPHA,A571,667;%%
\bibitem{GomezTejedor:1995pe}
  J.~A.~G\'omez Tejedor and E.~Oset,
  %``Double pion photoproduction on the nucleon: Study of the isospin
  %channels,''
  Nucl.\ Phys.\  A {\bf 600}, 413 (1996).
  %[arXiv:hep-ph/9506209].
  %%CITATION = NUPHA,A600,413;%%
%\cite{Bernard:1994ds}
\bibitem{Bernard:1994ds}
  V.~Bernard, N.~Kaiser, U.-G.~Mei{\ss}ner and A.~Schmidt,
  %``Threshold two pion photoproduction and electroproduction: More neutrals
  %than expected,''
  Nucl.\ Phys.\  A {\bf 580}, 475 (1994).
  %[arXiv:nucl-th/9403013].
  %%CITATION = NUPHA,A580,475;%%
%\cite{Bernard:1996ft}
\bibitem{Bernard:1996ft}
  V.~Bernard, N.~Kaiser and U.-G.~Mei{\ss}ner,
  %``Double neutral pion photoproduction at threshold,''
  Phys.\ Lett.\  B {\bf 382}, 19 (1996).
  %[arXiv:nucl-th/9604010].
  %%CITATION = PHLTA,B382,19;%%
%\cite{Bernard:1995tf}
\bibitem{Bernard:1995tf}
  V.~Bernard, U.-G.~Mei{\ss}ner and N.~Kaiser,
  %``Comment on 'Low-energy expansions for double pion photoproduction.',''
  Phys.\ Rev.\ Lett.\  {\bf 74}, 1036 (1995).
  %%CITATION = PRLTA,74,1036;%%
\bibitem{Nacher:2000eq}
  J.~C.~Nacher, E.~Oset, M.~J.~Vicente and L.~Roca,
  %``The role of $\Delta(1700)$ excitation and $\rho$ production in double  pion
  %photoproduction,''
  Nucl.\ Phys.\  A {\bf 695}, 295 (2001).
  %[arXiv:nucl-th/0012065].
  %%CITATION = NUPHA,A695,295;%%
\bibitem{Nacher:2001yr}
  J.~C.~Nacher and E.~Oset,
  %``Study of polarization observables in double pion photoproduction on the
  %proton,''
  Nucl.\ Phys.\  A {\bf 697}, 372 (2002).
  %[arXiv:nucl-th/0106005].
  %%CITATION = NUPHA,A697,372;%%
\bibitem{Mokeev:2001qg}
  V.~I.~Mokeev {\it et al.},
  %``Phenomenological Model For Describing Pion Pair Production On A Proton By
  %Virtual Photons In The Energy Region Of Nucleon Resonance Excitation,''
  Phys.\ Atom.\ Nucl.\  {\bf 64}, 1292 (2001)
  [Yad.\ Fiz.\  {\bf 64}, 1368 (2001)].
  %%CITATION = YAFIA,64,1368;%%
\bibitem{Roca:2004vs}
  L.~Roca,
  %``Helicity asymmetries in double pion photoproduction on the proton,''
  Nucl.\ Phys.\  A {\bf 748}, 192 (2005).
  %[arXiv:nucl-th/0407049].
  %%CITATION = NUPHA,A748,192;%%
\bibitem{Fix:2005if}
  A.~Fix and H.~Arenh\"ovel,
  %``Double pion photoproduction on nucleon and deuteron,''
  Eur.\ Phys.\ J.\  A {\bf 25}, 115 (2005).
  %[arXiv:nucl-th/0503042].
  %%CITATION = EPHJA,A25,115;%%
\bibitem{Roberts:2004mn}
  W.~Roberts and T.~Oed,
  %``Polarization observables for two-pion production off the nucleon,''
  Phys.\ Rev.\  C {\bf 71}, 055201 (2005).
  %[arXiv:nucl-th/0410012].
  %%CITATION = PHRVA,C71,055201;%%
\bibitem{Kamano:2009im}
  H.~Kamano, B.~Juli\'a D\'iaz, T.~S.~Lee, A.~Matsuyama and T.~Sato,
  %``Double and single pion photoproduction within a dynamical coupled-channels
  %model,''
  Phys.\ Rev.\  C {\bf 80}, 065203 (2009).
  %[arXiv:0909.1129 [nucl-th]].
  %%CITATION = PHRVA,C80,065203;%%
\bibitem{naka}
  T.~Nakabayashi {\it et al.},
  %``Photoproduction of eta mesons off protons for Egamma <= 1.15-GeV,''
  Phys.\ Rev.\  C {\bf 74}, 035202 (2006).
  %%CITATION = PHRVA,C74,035202;%%
\bibitem{ajaka}
  J.~Ajaka {\it et al.},
  %``Simultaneous photoproduction of eta and pi0 mesons on the proton,''
  Phys.\ Rev.\ Lett.\  {\bf 100}, 052003 (2008).
  %%CITATION = PRLTA,100,052003;%%
\bibitem{Horn:2007pp}
  I.~Horn {\it et al.}  [CB-ELSA Collaboration],
  %``Evidence for a parity doublet $\Delta(1920)P_{33}$ and $\Delta(1940)D_{33}$
  %from $\gamma p\to p\pi^0\eta$,''
  Phys.\ Rev.\ Lett.\  {\bf 101}, 202002 (2008).
  %[arXiv:0711.1138 [nucl-ex]].
  %%CITATION = PRLTA,101,202002;%%
\bibitem{Horn:2008qv}
  I.~Horn {\it et al.}  [CB-ELSA Collaboration],
  %``Study of the reaction $\gamma p\to p\pi^0\eta$,''
  Eur.\ Phys.\ J.\  A {\bf 38}, 173 (2008).
  %[arXiv:0806.4251 [nucl-ex]].
  %%CITATION = EPHJA,A38,173;%%
\bibitem{Kashevarov:2009ww}
  V.~L.~Kashevarov,
  %``Photoproduction of pi^0 eta on protons and the Delta(1700)D_{33}
  %resonance,''
  Eur.\ Phys.\ J.\  A {\bf 42}, 141 (2009).
  %[arXiv:0901.3888 [hep-ex]].
  %%CITATION = EPHJA,A42,141;%%
\bibitem{Gutz:2008zz}
  E.~Gutz {\it et al.}  [CBELSA/TAPS Collaboration],
  %``Measurement of the beam asymmetry Sigma in pi eta: Production off the
  %proton with the CBELSA/TAPS experiment,''
  Eur.\ Phys.\ J.\  A {\bf 35}, 291 (2008).
  %%CITATION = EPHJA,A35,291;%%
\bibitem{Gutz:2009zh}
  E.~Gutz {\it et al.} [CBELSA/TAPS Collaboration],
  %``Photoproduction of meson pairs: First measurement of the polarization
  %observable I^s,''
  Phys.\ Lett.\  B {\bf 687}, 11 (2010).  
  %[arXiv:0912.2632 [nucl-ex]].
  %%CITATION = ARXIV:0912.2632;%%
\bibitem{Kashevarov:2010gk}
  V.~L.~Kashevarov {\it et al.}  [A2 Collaboration],
  %``First measurement of the circular beam asymmetry in the gamma p --> pi0 eta
  %p reaction,''
  arXiv: 1009.4093 [nucl-ex].
  %%CITATION = ARXIV:1009.4093;%%
\bibitem{Jido:2001nt}
  D.~Jido, M.~Oka and A.~Hosaka,
  %``Chiral symmetry of baryons,''
  Prog.\ Theor.\ Phys.\  {\bf 106}, 873 (2001).
  %[arXiv:hep-ph/0110005].
  %%CITATION = PTPKA,106,873;%%
\bibitem{Doring:2005bx}
  M.~D\"oring, E.~Oset and D.~Strottman,
  %``Chiral dynamics in the gamma p --> pi0 eta p and gamma p --> pi0 K0  Sigma+
  %reactions,''
  Phys.\ Rev.\  C {\bf 73}, 045209 (2006).
  %[arXiv:nucl-th/0510015].
  %%CITATION = PHRVA,C73,045209;%%
\bibitem{mariana} M. Nanova and V. Metag, private communication.
\bibitem{Fix:2008pv}
  A.~Fix, M.~Ostrick and L.~Tiator,
  %``Analysis of angular distributions in gamma N \to pi0 eta N,''
  Eur.\ Phys.\ J.\  A {\bf 36}, 61 (2008).
  %[arXiv:0803.1755 [nucl-th]].
  %%CITATION = EPHJA,A36,61;%%
\bibitem{Anisovich:2004zz}
  A.~Anisovich, E.~Klempt, A.~Sarantsev and U.~Thoma,
  %``Partial wave decomposition of pion and photoproduction amplitudes,''
  Eur.\ Phys.\ J.\  A {\bf 24}, 111 (2005).
  %[arXiv:hep-ph/0407211].
  %%CITATION = EPHJA,A24,111;%%
\bibitem{Anisovich:2007bq}
  A.~V.~Anisovich, V.~Kleber, E.~Klempt, V.~A.~Nikonov, A.~V.~Sarantsev and U.~Thoma,
  %``Baryon resonances and polarization transfer in hyperon photoproduction,''
  Eur.\ Phys.\ J.\  A {\bf 34}, 243 (2007).
  %[arXiv:0707.3596 [hep-ph]].
  %%CITATION = EPHJA,A34,243;%%
\bibitem{Kolomeitsev:2003kt}
  E.~E.~Kolomeitsev and M.~F.~M.~Lutz,
  %``On baryon resonances and chiral symmetry,''
  Phys.\ Lett.\  B {\bf 585}, 243 (2004).
  %[arXiv:nucl-th/0305101].
  %%CITATION = PHLTA,B585,243;%% 
\bibitem{Sarkar:2004jh}
  S.~Sarkar, E.~Oset and M.~J.~Vicente Vacas,
  %``Baryonic resonances from baryon decuplet-meson octet interaction,''
  Nucl.\ Phys.\  A {\bf 750}, 294 (2005)
  [Erratum-ibid.\  A {\bf 780}, 78 (2006)].
  %[arXiv:nucl-th/0407025].
  %%CITATION = NUPHA,A750,294;%%
\bibitem{Doring:2007rz}
  M.~D\"oring,
  %``Radiative decay of the Delta(1700),''
  Nucl.\ Phys.\  A {\bf 786}, 164 (2007).
  %[arXiv:nucl-th/0701070].
  %%CITATION = NUPHA,A786,164;%%
\bibitem{Amsler:2008zzb}
  C.~Amsler {\it et al.}  [Particle Data Group],
  %``Review of particle physics,''
  Phys.\ Lett.\  B {\bf 667}, 1 (2008).
  %%CITATION = PHLTA,B667,1;%%
\bibitem{Doring:2006pt}
  M.~D\"oring, E.~Oset and D.~Strottman,
  %``The pi- p --> K0 pi0 Lambda reaction and clues to the nature of the
  %Delta*(1700) resonance,''
  Phys.\ Lett.\  B {\bf 639}, 59 (2006).
  %[arXiv:nucl-th/0602055].
  %%CITATION = PHLTA,B639,59;%%
\bibitem{Nanova:2008kr}
  M.~Nanova {\it et al.}  [CBELSA/TAPS Collaboration],
  %``K^0 pi^0 Sigma^+ and K^*0 Sigma^+ photoproduction off the proton,''
  Eur.\ Phys.\ J.\  A {\bf 35}, 333 (2008).
  %[arXiv:0803.2146 [nucl-ex]].
  %%CITATION = EPHJA,A35,333;%%
\bibitem{Fix:2010bd}
  A.~Fix, V.~L.~Kashevarov, A.~Lee and M.~Ostrick,
  %``Isobar model analysis of pi0-eta photoproduction on protons,''
  Phys.\ Rev.\  C {\bf 82}, 035207 (2010).
  %arXiv:1004.5240 [nucl-th].
  %%CITATION = ARXIV:1004.5240;%%
\bibitem{Borasoy:2005zg}
  B.~Borasoy, P.~C.~Bruns, U.-G.~Mei{\ss}ner and R.~Ni{\ss}ler,
  %``Gauge invariance in two-particle scattering,''
  Phys.\ Rev.\  C {\bf 72}, 065201 (2005).
  %[arXiv:hep-ph/0508307].
  %%CITATION = PHRVA,C72,065201;%%
\bibitem{Borasoy:2007ku}
  B.~Borasoy, P.~C.~Bruns, U.-G.~Mei{\ss}ner and R.~Ni{\ss}ler,
  %``A gauge invariant chiral unitary framework for kaon photo- and
  %electroproduction on the proton,''
  Eur.\ Phys.\ J.\  A {\bf 34}, 161 (2007).
  %[arXiv:0709.3181 [nucl-th]].
  %%CITATION = EPHJA,A34,161;%%
\bibitem{Nacher:1999ni}
  J.~C.~Nacher, E.~Oset, H.~Toki and A.~Ramos,
  %``Radiative production of the Lambda(1405) resonance in K- collisions on
  %protons and nuclei,''
  Phys.\ Lett.\  B {\bf 461}, 299 (1999).
  %[arXiv:nucl-th/9902071].
  %%CITATION = PHLTA,B461,299;%%
\bibitem{gutzprivate} E. Gutz, private communication.
\bibitem{gutzpreparation} CBELSA/TAPS, publication in preparation.
\bibitem{Arndt:2006bf}
  R.~A.~Arndt, W.~J.~Briscoe, I.~I.~Strakovsky and R.~L.~Workman,
  %``Extended Partial-Wave Analysis of piN Scattering Data,''
  Phys.\ Rev.\  C {\bf 74}, 045205 (2006).
  %[arXiv:nucl-th/0605082].
  %%CITATION = PHRVA,C74,045205;%%
\bibitem{Inoue:2001ip}
  T.~Inoue, E.~Oset and M.~J.~Vicente Vacas,
  %``Chiral unitary approach to S-wave meson baryon scattering in the
  %strangeness S=0 sector,''
  Phys.\ Rev.\  C {\bf 65}, 035204 (2002).
 % [arXiv:hep-ph/0110333].
  %%CITATION = PHRVA,C65,035204;%%
\bibitem{Doring:2009uc}
  M.~D\"oring and K.~Nakayama,
  %``The phase and pole structure of the N*(1535) in piN-->piN and
  %gammaN-->piN,''
  Eur.\ Phys.\ J.\  A {\bf 43}, 83 (2010).
  %[arXiv:0906.2949 [nucl-th]].
  %%CITATION = EPHJA,A43,83;%%
\bibitem{Doring:2009qr}
  M.~D\"oring and K.~Nakayama,
  %``On the cross section ratio sigma_n/sigma_p in eta photoproduction,''
  Phys.\ Lett.\  B {\bf 683}, 145 (2010).
  %arXiv:0909.3538 [nucl-th].
  %%CITATION = ARXIV:0909.3538;%%
\end{thebibliography}
\end{document}